\documentclass[a4paper,12pt,dvips,]{scrartcl}%{article}
\usepackage{graphicx}

\newcommand\beq{\begin{equation}}
\newcommand\eeq{\end{equation}}
\newcommand\bea{\begin{eqnarray}}
\newcommand\eea{\end{eqnarray}}
\begin{document}
\begin{titlepage} 
\vspace{0.2in}

\begin{center} {\LARGE \bf 
On the Quasi-Isotropic Inflationary Solution
} 
\vspace*{2cm}

{\bf Giovanni Imponente$~^{1,3,4}$  \\
Giovanni Montani$~^{2,4}$ } 

\vspace*{2cm}
$~^1$ Dipartimento di Fisica Universit\'a di Napoli ``Federico II'' \\
$~^2$ Dipartimento di Fisica Universit\'a di Roma ``La Sapienza'' \\
$~^3$ INFN --- Napoli \\ 
$~^4$ ICRA --- International Center for Relativistic Astrophysics \\ 
c/o Dip. Fisica (G9), Piazzale Aldo Moro 5, 00185 Roma, Italy\\
\vspace*{1cm}
e-mail: imponente@icra.it, montani@icra.it
%\vspace*{1cm}

\vspace{\stretch{1}}

\end{center} \indent


\begin{abstract}

In this paper we find a solution 
for a quasi-isotropic inflationary 
Universe which allows to introduce in the 
problem a certain degree of inhomogeneity. 
We consider a model which generalizes 
the (flat) FRW one by introducing a first 
order inhomogeneous term, whose 
dynamics is induced by an effective 
cosmological constant. 
The 3-metric tensor is constituted 
by a dominant term, corresponding to 
an isotropic-like component, while 
the amplitude of the first order one 
is controlled by a ``small''  
function $\eta(t)$. 

In a Universe filled with ultra relativistic
matter and a real self-interacting scalar 
field, we discuss the resulting dynamics, 
up to first order in $\eta$, when the scalar 
field performs a slow roll on 
a plateau of a symmetry breaking 
configuration and induces an effective 
cosmological constant.

We show how the spatial distribution 
of the ultra relativistic matter and of 
the scalar field admits an arbitrary 
form but nevertheless, due to the required
inflationary e-folding, 
it cannot play a serious dynamical
role in tracing the process of structures 
formation (via the Harrison--Zeldovic 
spectrum). As a consequence, this paper 
reinforces the idea
 that the inflationary 
scenario is incompatible with a classical
origin of the large scale structures.
\end{abstract}

\end{titlepage}

\section{General Statement}

The inflationary model \cite{G81}-\cite{CW73}  
is, up to now,
the most natural and complete scenario to make
account of the problems outstanding in the
Standard Cosmological Model, 
like the horizons and flatness 
paradoxes \cite{KT90} 
(for pioneer works on inflationary 
scenario and the spectrum of gravitational 
perturbation, see also \cite{STA80,STA79}); 
indeed such
a dynamical scheme, on one hand is able to justify
the high isotropy of the cosmic microwaves 
background radiation
(characterized by temperature fluctuations
$\mathcal{O}(10^{-4})$ \cite{dB01}) and, on the other one, 
provides a mechanism for generating 
a (scale invariant) 
spectrum of inhomogeneous perturbations
(via the scalar field quantum fluctuations).\\ 
Moreover, as shown in \cite{KM02,Star83}, a
slow-rolling phase of the scalar field allows 
to connect the generic inhomogeneous Mixmaster 
dynamics \cite{BKL70}-\cite{IM02}
with a later quasi-isotropic Universe
evolution (in principle 
compatible with the actual cosmological
 picture).\\
With respect to this,  we investigate the 
dynamics performed by small inhomogeneous
 corrections to a leading order metric, 
during inflationary expansion.

The model presented has the relevant
feature to contain inhomogeneous corrections
to a flat FRW Universe, which in principle 
could take a role to understand the process of 
structures formation, even in presence of an 
inflationary behavior;
however, a careful analysis of our result prevents
this possibility in view of the strong inflationary
e-folding, so confirming the expected incompatibility 
between an inflationary scenario and a classical origin
of the Universe clumpyness. 

In what follows, we will use the so-called 
{\em quasi isotropic solution} 
which was introduced in \cite{LK63} as the 
simplest, but rather general, extension of the 
FRW model; for a discussion of the quasi isotropic 
solution in the framework of the 
``long-wavelength'' approximation, see \cite{DT94}
while for the implementation of such solution
after inflation to generic equation of state
and to the case of two ideal hydrodynamic fluid
see, respectively, \cite{KHA02} and \cite{KHA03}. \\
In \cite{M99} (see also \cite{M00}) 
this solution is discussed in the presence
of a real scalar field kinetic energy, which 
leads to a power-law solution for the 3-metric, 
and predicts interesting features for the 
ultra relativistic matter dynamics. 

In the present paper we analyse
the opposite dynamical scheme, when the scalar 
field undergoes a slow-rolling phase since the 
effective cosmological constant dominates its 
kinetic energy.
We provide, 
up to the first two orders of approximation 
and in a synchronous reference,
a detailed description of
the 3-metric, of the scalar field
and of the ultra relativistic matter dynamics,
showing that the
volume of the Universe expands exponentially 
and induces a corresponding exponential decay 
(as the inverse fourth power of the cosmic 
scale factor), either of the 3-metric 
corrections, as well as of the 
ultra relativistic matter
(the same behavior characterizes roughly even 
the scalar field inhomogeneities).
It is remarkable that the spatial dependence 
of such component is described by a function 
which remains an arbitrary degree of freedom; 
in spite of such freedom in fixing the primordial 
spectrum of inhomogeneities, 
due to the inflationary e-folding,
we show there is no chance that,
after the de-Sitter phase,
such relic perturbations can survive enough 
to trace the large scale
structures formation by an 
Harrison--Zeldovic spectrum.\\
This behavior suggests that the spectrum of
inhomogeneous perturbations \cite{MB95} cannot 
arise, in such a model, directly by the 
classical field nature, instead of by its 
quantum dynamics.

Finally, we recall that the presence of 
the kinetic term of a scalar field,
here regarded as negligible, 
induces,
near enough to the singularity, 
a deep modification of the general cosmological 
solution, leading to the appearance of a dynamical 
regime, during which, point by point in space, 
the three spatial directions 
behaves monotonically \cite{BK73,B99}. 

In Section 2 is presented a brief review 
concerning the origin of spectrum perturbations
in the inflationary scenario.
We review in Section 3
the general formulation of the 
cosmological problem corresponding 
to the dynamics of a tridimensional 
Universe filled of ultra relativistic matter, 
and in which lives a real self-interacting 
scalar field. 
In Section 4 it is introduced the 
quasi--isotropic formalism, which is 
applied in Section 5 to get, 
far enough from the singularity,
the inflationary solution; 
finally, in Section 6 follow
some concluding remarks and physical 
considerations on the model obtained.

%%%%%%%%%%%%%%%%%%%%%%%%%%%%%%%%%%%%%inizio parte inserita luglio2003

\section{Inhomogeneous Perturbations 
from an Inflationary Scenario}

The theory of inflation is based on the idea that during the
Universe evolution takes place a phase transition
(for instance associated with a spontaneous symmetry breaking of a
Grand-Unification model of strong and electroweak interactions)
which induces an effective cosmological constant dominating the
expansion dynamics. As a result arises
an exponential expansion of the
Universe and, under a suitable fine-tuning of the parameters,
it is able to ``stretch'' so strongly the geometry that 
two fundamental and puzzling problems of
the Standard Cosmological Model, like
the  ones of the so-called {\em horizons} and
{\em flatness paradoxes}
are naturally solved.\\
In the ``new inflation'' theory, the Universe
undergoes a de-Sitter phase
when the scalar field performs a ``slow-rolling'' behavior
over a very flat region of the potential
between the false and true vacuum.
Such an exponential expansion ends with the
scalar field falling down in the potential well
associated to the real vacuum. Here the scalar field dies
via damped
(by the expansion of the Universe and particles creation) 
oscillations which reheat the cold Universe left by the
de-Sitter expansion (we recall that the relativistic particles
temperature is proportional to the inverse scale factor). 
Indeed, the decay of this super-cooled bosons condensate
into relativistic particles, being a typical irreversible process,
generates a huge amount of entropy, which allows to account for
the present high value ($\sim \mathcal{O}(10^{88})$)
of the Universe entropy per
comoving volume.

Apart from the transition across the
potential barrier between false and true vacuum,
which takes place in general via a tunneling, the whole
inflationary dynamics can be satisfactorily described via
a classical uniform scalar field $\phi = \phi (t)$.
The assumption that the field behaves in a classical way
is supported by its bosonic and cosmological nature, but the
existence of quantum fluctuations of the field within the different
inflationary ``bubbles'' leads to relax the hypothesis of
dealing with a perfectly uniform scalar field.

In general, when analyzing density perturbations,
turns out convenient to introduce the dimensionless
quantity

\begin{equation}
\delta\rho(t, x^{\gamma })\equiv 
\frac{\Delta\rho(t, x^{\gamma })}{\bar\rho}=
\frac{\rho - \bar\rho}{\bar\rho}
\, , 
\label{xax}
\end{equation}

where $\bar\rho$ denotes the mean density 
and $\gamma=1,2,3$.
The best formulation of the density perturbations theory           
is obtained expanding $\delta \rho$ 
in its Fourier components, or modes, 

\begin{equation}
\delta\rho_k=\frac{1}{(2\pi )^{3}}   \int d^3x \,e^{ik_{\alpha }x^{\alpha }}
\delta\rho(t, x^{\gamma })
\, .
\label{xax1}
\end{equation}

As long as the perturbations are in
the linear regime, i.e. $\delta\rho_k\ll 1$,
we can follow appropriately the dynamics 
of each mode with wavenumber 
$k$, which corresponds to a wavelength
$\lambda=\frac{2\pi}{k}$;
however, in an expanding Universe, the
\emph{physical} size
of the perturbations evolves via the \emph{cosmic scale factor}
$a(t)$,

Since in the Standard Cosmological Model,
the ``Hubble radius''
scales as $H^{-1}\propto t$, while $a(t)\propto t^n$
with $n<1$, then every
perturbation, now inside the Hubble radius, 
was outside it at some earlier time.
We stress how 
the perturbations with a physical size, respectively
smaller or greater than the Hubble radius,
have a very different dynamics, the former ones 
being affected by the action of the microphysics processes.

In the case of an inflationary scenario,
the situation is quite different. Since during the de-Sitter phase 
the Hubble radius remains constant, while the cosmic scale
factor ``explodes'' exponentially; 
hence, all cosmological interesting scales have crossed the horizon
twice, i.e.
the perturbations begin sub-horizon sized,
cross the Hubble radius
during inflation and later cross back again inside
the horizon. \\
This feature has a strong implication on the
initial spectrum of density perturbations
predicted by inflation.
We present a qualitative argument to understand
how this spectrum 
can be generated. 

During inflation, the density perturbations
are expected to arise from the quantum mechanical
fluctuations of the scalar field $\phi$; 
these are, as usual, decomposed in their Fourier components
$\delta\phi_k$, i.e.

\begin{equation}
\delta\phi_k=\frac{1}{(2\pi )^{3}} \int d^3x\,e^{ik_{\alpha }x^{\alpha }}
\delta\phi (t, x^{\gamma })
\, , 
\label{xax2}
\end{equation}

The spectrum of quantum mechanical fluctuations of the
scalar field is defined as
\begin{equation}
(\Delta\phi)^2_k\equiv
\frac1{\cal V}\frac{k^3}{2\pi^2}\left|\delta\phi_k\right|^2
\label{xax3}
\, ,
\end{equation}
where ${\cal V}$ denotes the comoving volume.
For a massless minimally coupled scalar field in a de-Sitter
space-time,
which approximates very well the real physical situation during
the Universe exponential expansion, it is well known that
(see \cite{KT90} \S 8.4 and references therein)

\begin{equation}
(\Delta\phi)^2_k=\left(\frac{H}{2\pi}\right)^2
\label{xax4}
\, ,
\end{equation}
then the mean square fluctuation of $\phi$,
$(\Delta\phi)^2$, takes the following form:

 \begin{equation}
(\Delta\phi)^2=\frac1{(2\pi)^3V}\int d^3k\left|\delta\phi_k\right|^2=
\int\left(\frac{H}{2\pi}\right)^2d(\ln k)
\label{xax5}
\, .
\end{equation}

Since $H$ is constant during the de-Sitter phase of
the Universe, 
each mode $k$ contributes roughly
the same amplitude to the mean square fluctuation. 
Indeed, 
the only dependence on $k$ takes place in 
the logarithmic term, but  
the modes of cosmological interest lay 
between $1~ Mpc$ and $3000~ Mpc$
(it is commonly adopted the convention to set
the actual cosmic scale factor equal to unity), 
corresponding to a logarithmic interval
of less than an order of magnitude.

Thus we can conclude that
any mode $k$ crosses the horizon
having almost a constant amplitude
$\delta\phi_k\simeq H/2\pi$.
A delicate question concerns the mechanism by which
such quantum fluctuations of the scalar field achieve
a classical nature(for a detailed 
discussion see \cite{STA96}); 
here we simply observe how
each mode $k$, once reached a classical stage,
is subjected to the dynamics

\begin{equation}
 \delta\ddot \phi_k+3H\delta\dot\phi_k +
 \frac{ k^2}{a^2} {\delta\phi_k}=0\label{mot}
\, ; 
\label{xax6}
\end{equation}
according to this equation, 
 super-horizon modes $k\ll aH$
(i.e. $\lambda _{phys}\gg H^{-1}$) 
admit the trivial dynamics (\ref{mot}) with
$\delta\phi_k\sim \mathrm{const.}$.
This simple analysis implies the important feature 
that any mode re-enters the horizon with
roughly the same amplitude it had at the first
horizon crossing. 
The spectrum of perturbations so generated is
then induced into the relativistic energy density
coming from the reheating phase, associated with the
bosons decay; since that moment the evolution of
the perturbation spectrum follows a standard
paradigm.

Finally, we discuss another feature of the
quantum mechanical fluctuations
generated by the inflation, regarding 
their gaussian distribution: as long as
the field $\phi$ is minimally coupled, it has a
low self interaction and
each mode fluctuates independently; hence, 
since the fluctuations we actually observe are
the sum of many of its quantum
ones, their distribution is expectable 
to be gaussian (as it should
be for the sum of many independent variables).

%%%%%%%%%%%%%%%%%%%%%%%%%%%%%%%%%%%%%fine parte inserita luglio2003

\section{Field Equations in the Synchronous Reference}

The line element
in a synchronous reference frame of 
coordinates $(t, x^\gamma)$ (we adopt 
units in which the speed of 
light $c$ is equal to unity)\footnote{Greek indexes 
$\alpha,\beta,\gamma$ take values 1,2,3 labeling
the spatial coordinates on the space-like hypersurfaces 
of constant proper cosmological
time $t$, while Latin ones label 0 to 3.}
writes 
\beq
ds^2 = dt^2 - \gamma_{\alpha\beta}(t, x^\gamma)
dx^{\alpha }dx^{\beta } \, ,
\label{a}
\eeq
where ${\gamma }_{\alpha \beta }(t, x^{\gamma })$ 
is the three-dimensional 
metric tensor describing the geometry of the spatial slices. 

Let us describe the matter by a perfect fluid with  
ultra relativistic equation of state 
$p = \frac{\epsilon }{3}$ ($p$ and $\epsilon $ 
denote respectively the fluid pressure and energy 
density) and a scalar field $\phi (t, x^{\gamma })$ 
with a potential term $V(\phi )$. \\
In what follows, we write the Einstein 
equations as
\beq
\label{ee}
R_i^k = \chi \sum_{(z)=m,\phi}{\left({T_i^k}^{(z)} -
 \frac{1}{2}\delta_i^k {T_l^l}^{(z)} \right)}  
\eeq
where $\chi $ denotes the Einstein constant 
$\chi = 8\pi G$ ($G$ being the Newton constant) and 
${T_i^k}^{(m)}$ and ${T_i^k}^{(\phi)}$ indicate, 
respectively, the energy-momentum tensor of
the matter and the scalar field.
Explicitly, in a synchronous reference, such 
equations reduce to the
partial differential system 
\bea 
\frac{1}{2} \partial _t k_{\alpha }^{\alpha } + \frac{1}{4}
k_{\alpha }^{\beta }k_{\beta }^{\alpha } = 
\chi \left[ - (4{u_0}^2 - 1)\frac{\epsilon }{3} -  
(\partial _t\phi )^2 + V(\phi )\right] 
\label{b}  \\
\frac{1}{2}(k^{\beta }_{\alpha ;\beta } - 
k^{\beta }_{\beta ;\alpha }) = 
\chi \left( \frac{4}{3}\epsilon u_{\alpha }u_0 +
\partial _{\alpha }\phi \partial _t \phi \right) 
\label{c} \\
\frac{1}{2\sqrt{\gamma }}\partial _t (\sqrt{\gamma }
k_{\alpha }^{\beta }) + P_{\alpha }^{\beta } = 
\chi \left[ {\gamma }^{\beta \gamma }\left( 
\frac{4}{3}\epsilon u_{\alpha }u_{\gamma } 
+ \partial _{\alpha }\phi \partial _{\gamma }\phi \right) + 
\left( \frac{\epsilon }{3} 
+ V(\phi )\right) {\delta }_{\alpha }^{\beta }
\right] \, ,
\label{d} 
\eea
where the vector field $u_i$  
($i=0,\ldots,3$) represents the matter four-velocity
and we used the notations 
\bea
\partial _t(\quad )\equiv \frac{\partial(\quad ) }{\partial t } 
\, ,  \qquad \partial _{\alpha }(\quad ) \equiv 
\frac{\partial (\quad )}{\partial x^{\alpha }} \, , 
\label{e} \\
\gamma \equiv det \gamma_{\alpha \beta} \, ,\qquad 
k_{\alpha \beta } \equiv \partial _t{\gamma }_{\alpha \beta } 
\, ,  \qquad k_{\alpha }^{\beta } = 
{\gamma }^{\beta \gamma }k_{\alpha \gamma } \, .
\label{f}
\eea
The metric ${\gamma }_{\alpha \beta}$ allows  
to construct the three-dimensional Ricci tensor 
$P_{\alpha }^{\beta } = 
{\gamma }^{\beta \gamma }P_{\alpha \gamma }$ as
\beq
P_{\alpha \beta} = 
\partial _{\gamma }{\lambda }^{\gamma }_{\alpha \beta } 
- \partial _{\alpha }{\lambda }^{\gamma }_{\beta \gamma } + 
{\lambda }^{\gamma }_{\alpha \beta }{\lambda }^{\delta }_{\gamma \delta } 
- {\lambda }^{\gamma }_{\alpha \delta }{\lambda }^{\delta }_{\beta \gamma } 
\label{g}
\eeq
in which appear the pure spatial Christoffel symbols 
\beq
{\lambda }^{\gamma }_{\alpha \beta } \equiv  
\frac{1}{2}{\gamma }^{\gamma 
\delta }(\partial _{\alpha }{\gamma }_{\delta \beta } + 
\partial _{\beta }{\gamma }_{\alpha \delta } 
-\partial _{\delta }{\gamma }_{\alpha \beta }) 
\label{h} 
\label{i}
\eeq 
also used to form the covariant
derivative $(\quad )_{;\alpha }$. 

The dynamics of the scalar field 
$\phi (t, x^{\gamma })$ 
is described by a partial differential 
equation, coupled to the above Einsteinian 
system, which in a synchronous reference reads  
\beq 
\partial _{tt}\phi + \frac{1}{2}k_{\alpha }^{\alpha }
\partial _t\phi - {\gamma }^{\alpha \beta }{\phi }_{;\alpha ;\beta } 
+ \frac{dV}{d\phi } = 0 
\label{m} 
\eeq
where we adopted the obvious notation 
\beq 
\partial _{tt} (\quad ) \equiv \frac{\partial ^2(\quad )}
{\partial t^2}  \, .
\label{n} 
\eeq
The hydrodynamic equations,
taking into account for the matter 
evolution,
in a synchronous reference and for the 
ultra relativistic case, 
possess the structure \cite{LK63}
\bea 
\frac{1}{\sqrt{\gamma }}\partial _t(\sqrt{\gamma }{\epsilon }^{3/4}u_0) + 
\frac{1}{\sqrt{\gamma }}
\partial _{\alpha }(\sqrt{\gamma }{\epsilon }^{3/4}u^{\alpha }) = 0 
\label{o} \\ 
4{\epsilon }\left( \frac{1}{2}\partial _t{u_0}^2 + u^{\alpha }
\partial _{\alpha }u_0 + \frac{1}{2}k_{\alpha \beta }u^{\alpha }
u^{\beta }\right) = (1 - {u_0}^2)\partial _t\epsilon - u_0u^{\alpha }
\partial _{\alpha }\epsilon 
\label{p}\\ 
4\epsilon \left( u_0\partial _tu_{\alpha } + u^{\beta }
\partial _{\beta }u_{\alpha } + \frac{1}{2}u^{\beta }u^{\gamma }
\partial  _{\alpha }{\gamma }_{\beta \gamma }\right) = 
- u_{\alpha }u_0\partial _t\epsilon + 
({\delta }_{\alpha }^{\beta }  - 
u_{\alpha }u^{\beta })\partial _{\beta }\epsilon \, .
\label{q}
\eea 
In view of the chosen feature for (\ref{ee}), 
equation (\ref{o}) doesn't contain spatial 
gradients of the 3-metric tensor and of the 
scalar field.
This scheme is completed by observing how it be 
covariant with respect to coordinates 
transformation of the form 
\beq 
t^{\prime } = t + f(x^{\gamma }) \,, \qquad  
x^{\alpha \prime } = x^{\alpha \prime }(x^{\gamma }) 
\label{r} 
\eeq
being $f$ a generic space dependent function.

\section{Quasi isotropic Model}

In order to introduce in a quasi isotropic (inflationary) 
scenario small inhomogeneous 
corrections to the leading order, we require a 
tridimensional metric tensor having the following structure 
\bea
{\gamma }_{\alpha \beta }(t, x^{\gamma }) &=& 
a^2(t){\xi }_{\alpha \beta }(x^{\gamma }) + 
b^2(t){\theta }_{\alpha \beta }(x^{\gamma }) 
+ \mathcal{O}(b^2) = \nonumber \\
&=& a^2(t)\left[ {\xi }_{\alpha \beta }(x^{\gamma }) + 
\eta (t){\theta }_{\alpha \beta }(x^{\gamma }) 
+ \mathcal{O}(\eta^2 )\right] 
\label{s} 
\eea
where we defined $\eta \equiv \frac{b^2}{a^2}$  
and suppose that $\eta$ satisfies the 
condition 
\beq 
\lim_{t\rightarrow \infty} \eta (t) = 0 \,.
\label{t}
\eeq 
We shall analyse the field equations 
(\ref{b})-(\ref{d}) retaining only 
terms linear in $\eta$ and its time 
derivatives. In the limit of the considered 
approximation, the inverse three-metric reads 
\beq
{\gamma }^{\alpha \beta }(t, x^{\gamma }) = 
\frac{1}{a^2(t)}\left( {\xi }^{\alpha \beta }(x^{\gamma }) -  
\eta (t){\theta }^{\alpha \beta }(x^{\gamma }) 
+ \mathcal{O}(\eta^2 )\right)  \, ,
\label{ss}
\eeq
where ${\xi }^{\alpha \beta }$ denotes the 
inverse matrix of ${\xi }_{\alpha \beta }$ 
and assumes a metric role, i.e. we have 
\beq 
{\xi }^{\beta \gamma }{\xi }_{\alpha \gamma } = 
{\delta }_{\alpha }^{\beta } \, \quad 
{\theta }^{\alpha \beta } = 
{\xi }^{\alpha \gamma }{\xi }^{\beta \delta }{\theta }_{\gamma \delta } \, .
\label{tt} 
\eeq
The covariant and contravariant 
three-metric expressions lead to the 
important explicit relations
\beq
k_{\alpha }^{\beta } = 2\frac{\dot{a}}{a}{\delta }_{\alpha }^{\beta } + 
\dot{\eta }{\theta }_{\alpha }^{\beta } \quad \Rightarrow \quad 
k_{\alpha }^{\alpha } = 6\frac{\dot{a}}{a} + \dot{\eta }\theta 
\, \quad \theta \equiv {\theta }_{\alpha }^{\alpha } 
\label{u} 
\eeq
where we set $(\quad )^. \equiv d(\quad )/dt$. 

Since it should take place the fundamental equality 
$\partial_t(\ln \gamma ) = k_{\alpha }^{\alpha }$, 
then we immediately get
\beq 
\gamma = j
a^6e^{\eta \theta } \quad \Rightarrow \quad \sqrt{\gamma } = 
\sqrt{j}a^3e^{\frac{1}{2}\eta \theta }\sim \sqrt{j}
a^3\left( 1 + \frac{1}{2}\eta \theta + \mathcal{O}(\eta^2 )\right) \,, \qquad  
j\equiv det{\xi }_{\alpha \beta } \, .
\label{v} 
\eeq 

Equations (\ref{b})-(\ref{d}) are analysed 
via the standard procedure of 
constructing asymptotic solutions in the limit 
$t\rightarrow \infty$, by verifying {\it a posteriori} 
the self-consistency of the approximation scheme, 
i.e. that the neglected 
terms were really of higher order in time.

\section{Inflationary Solution}

In the quasi-isotropic approach,
under consideration,
we assume that the scalar field dynamics, 
in the plateau region, be governed by a 
potential term as
\beq
V(\phi) = \Lambda + K(\phi) \, , 
\qquad \Lambda={\rm const.}
\label{w}
\eeq
where $\Lambda$ is the dominant term
and $K(\phi)$ is a small correction to 
it. The role of $K$, as shown in the 
following, is to contain
inhomogeneous 
corrections via the $\phi$-dependence;
the functional form of $K$ can
be any one of the most common inflationary 
potentials, 
as they appear near the flat 
region for the evolution of $\phi$.\\
What follows remains valid, for example, 
for the 
relevant cases of the quartic and 
Coleman--Weinberg 
expressions
\beq
\label{pot}
K(\phi)=\left\{
                \begin{array}{l} \displaystyle -\frac{\lambda}{4} \phi^4 \, ,  \qquad\qquad \qquad \lambda={\rm const.} \\
                \displaystyle B\phi^4\left[\ln\left(\frac{\phi^2}{\sigma^2}\right)-\frac{1}{2} \right] \, , \qquad \sigma={\rm const.} \, ,
                \end{array}
                \right. 
\eeq
viewed as corrections to the constant $\Lambda$ 
term, though explicit calculation are below 
developed only for the first case. 

Our inflationary solution is obtained under the 
standard requirements
\bea
\label{infla1}
\frac{1}{2}\left(\partial_t \phi\right)^2 \ll 
V\left(\phi\right) \\
\mid \partial_{tt}\phi \mid \ll \mid k^{\alpha}_{\alpha}\partial_t\phi \mid
\label{infla2} \, .
\eea
The above approximations and the substitution of 
(\ref{u}) reduce the scalar field 
equation (\ref{m}) to the form 
\beq
\left( 3\frac{\dot{a}}{a} + \frac{1}{2}\dot{\eta }\theta \right) 
\partial _t\phi - \lambda \phi^3= 0 
\label{x} 
\eeq
where we assumed that the contribution 
of the $\phi$ spatial gradients 
be negligible.

Similarly, the quasi-isotropic approach 
(in which the inhomogeneities become relevant
only for the next approximation order), 
once neglecting the spatial derivatives, 
in (\ref{o}), leads to 
\beq 
\sqrt{\gamma }{\epsilon }^{3/4}u_0 = 
l(x^{\gamma }) \quad 
\Rightarrow \quad \epsilon \sim \frac{l^{4/3}}{j^{2/3}a^4{u_0}^{4/3}} 
\left( 1 - \frac{2}{3}\eta \theta + \mathcal{O}(\eta^2 )\right) 
\label{a1} 
\eeq 
where $l(x^{\gamma })$ denotes an arbitrary 
function of the 
spatial coordinates. 

Let us now face,
in the same approximation scheme, 
the analysis of the Einstein
equations (\ref{b})-(\ref{d}). 
Taking into account (\ref{infla1}),
up to the first order in $\eta$, 
equation (\ref{b}) reads  
\beq
3\frac{\ddot{a}}{a} + 
\left[ \frac{1}{2}\ddot{\eta } + \frac{\dot{a}}{a}\dot{\eta } \right]
\theta - \chi \Lambda= - \chi 
\frac{\epsilon }{3}(3 + 4u^2) 
\label{a2} 
\eeq
having set 
\beq 
u^2 \equiv \frac{1}{a^2}{\xi }^{\alpha \beta }u_{\alpha }u_{\beta } 
\quad \Rightarrow \quad u_0 = \sqrt{1 + u^2}  \, .
\label{a3} 
\eeq
Equation (\ref{d}) reduces to the form 
\bea 
\frac{2}{3}(a^3)^{..}{\delta }_{\alpha }^{\beta } + 
(a^3\dot{\eta })^.{\theta }_{\alpha }^{\beta } + 
\frac{1}{3}[(a^3)^.\eta ]^.\theta {\delta }_{\alpha }^{\beta } +
aA_{\alpha}^{\beta}= \nonumber \\
= \chi\left[\frac{1}{a^2}\left( {\xi }^{\beta \gamma } -
\eta {\theta}^{\beta \gamma }\right)
\frac{4}{3}\epsilon u_{\alpha }u_{\gamma }
+ \left(\frac{\epsilon}{3} 
+\Lambda \right){\delta }_{\alpha }^{\beta }\right] 
2a^3\left( 1 + \frac{\eta \theta}{2} \right)  \, .
\label{a4} 
\eea
In this expression, the spatial curvature term  
reads, in the leading order, as 
\beq
P_{\alpha }^{\beta }(t, x^{\gamma }) = \frac{1}{a^2(t)}
A_{\alpha }^{\beta }(x^{\gamma }) \, ,
\label{a5} 
\eeq 
where $A_{\alpha \beta }(x^{\gamma }) = {\xi }_{\beta \gamma}
A_{\alpha }^{\gamma }$ denotes the Ricci tensor corresponding 
to ${\xi }_{\alpha \beta }(x^{\gamma })$. 

The trace of (\ref{a4}) gives the additional 
relation
\beq 
2(a^3)^{..} + 
(a^3{\eta})^{..}\theta + aA^\alpha_\alpha= 
\chi  
\left[ \frac{\epsilon}{3} \left( 3 + 4u^2 \right) 
+3\Lambda  \right] 
2a^3\left( 1 + \frac{\eta \theta}{2} \right) \, .
\label{a6} 
\eeq 
Comparing (\ref{a2}) with the trace (\ref{a6}), 
via their common term $(3+4u^2)\epsilon/3$, 
and estimating the different orders of magnitude, 
we get the following equations 
\bea 
(a^3)^{..} + 3a^2\ddot{a}- 4\chi a^3 \Lambda = 0 
\label{a7a} \\
A_{\alpha\beta}=0  
\label{a7b}\\
3(a^3\eta )^{..} + 3a^3\ddot{\eta } + 
2(a^3)^.\dot{\eta } + 9a^2\ddot{a}\eta - 
12 \chi a^3 \Lambda \eta = 0 \, .
\label{a7c} 
\eea 

Since (\ref{a7b}) implies that the
the tridimensional Ricci tensor
vanishes, and this 
condition corresponds to the vanishing of 
the Riemann tensor too, then we can conclude that
the obtained Universe is flat 
up to the leading order, i.e. 
\beq
\label{met}
\xi_{\alpha \beta}=\delta_{\alpha \beta} \, 
\qquad \Rightarrow \qquad j=1.
\eeq
Equation (\ref{a7a}) admits 
the expanding solution
\beq
a(t)=a_0 \exp\left\{ \frac{\sqrt{3\chi\Lambda}}{3}t\right\}
\label{a9}
\eeq
being $a_0$, 
the initial value
of the scale factor
amplitude, taken at the instant $t=0$ 
when the de-Sitter phase starts. 

Expression (\ref{a9}) for $a(t)$, when 
substituted in (\ref{a7c}) yields the 
differential equation for $\eta$ 
\beq
\ddot{\eta}+ \frac{4}{3}\sqrt{3\chi \Lambda}~\dot{\eta}=0 \, ,
\label{b2}
\eeq
whose only solution, satisfying the
limit (\ref{t}), reads 
\beq
\eta(t)=\eta_0 \exp\left\{-\frac{4}{3}\sqrt{3\chi\Lambda}~ t\right\} 
\qquad \Rightarrow \qquad \eta= \eta_0 \left( \frac{a_0}{a}\right)^4 \, ,
\label{b3}
\eeq
and, of course, we require $\eta_0 \ll a_0$.

Equations (\ref{a1}) and 
(\ref{a2}), in view of the solutions (\ref{a9}) 
for $a(t)$ and (\ref{b3}) for $\eta(t)$, 
are matched with consistency, by posing 
\bea 
u_{\alpha }(t, x^{\gamma }) = v_{\alpha }(x^{\gamma })
+ \mathcal{O}\left(\eta^2\right) \nonumber \\
(u_0)^2 = 1+ \mathcal{O}\left(\frac{1}{a^2}\right) 
\approx 1 \, , 
\label{b4}
\eea
and respectively 
\beq
\epsilon = - \frac{4}{3} \Lambda\eta \theta \, ,
\label{b5}
\eeq
which implies $\theta < 0$ for each point 
of the allowed domain of the spatial 
coordinates.
The comparison between (\ref{a1}) 
and (\ref{b5}) leads to the explicit 
expression also for $l(x^\gamma)$ in 
terms of $\theta$
\beq
\label{elle}
l(x^\gamma)=
\left(\frac{4}{3}\Lambda \eta_0{a_0}^4\right)^{3/4}
\left(-\theta\right)^{3/4} \, .
\eeq
Defining the auxiliary tensor with 
unit trace
${\Theta }_{\alpha \beta }(x^{\gamma })\equiv
{\theta }_{\alpha \beta }/\theta$,
the above analysis permits, from (\ref{a4}), 
to obtain for it the expression
\beq
{\Theta }_{\alpha}^{ \beta }= 
\frac{\delta_{\alpha}^{\beta}}{3} \, .
\label{b6}
\eeq

By (\ref{x}), the explicit 
form for $a$, once expanded in $\eta$, 
yields the first two
leading orders
of approximation for the scalar field
\beq
\phi\left(t, x^\gamma\right)= 
{\cal{C}}\sqrt{\frac{t_r}{t_r-t}}
\left(1- \frac{ 1}{4\sqrt{3\chi\Lambda}}
\frac{\eta}{t_r-t}\theta\right) \, ,
\qquad t_r =\frac{\sqrt{3\chi\Lambda}}{{\cal{C}}^22\lambda } \, ,
\label{b7}
\eeq
where ${\cal{C}}$ is an integration constant; 
finally, equation (\ref{d}) provides $v_{\alpha }$ 
in terms of $\theta$  
\beq
v_{\alpha }=-\frac{3}{4}\frac{1}{\sqrt{3\chi\Lambda}}
\partial_\alpha\ln\mid\theta\mid  \, .
\label{b8}
\eeq 

On the basis of (\ref{b6})-(\ref{b8}),
the hydrodynamic equations (\ref{o})-(\ref{q}) 
reduce to an identity, in the leading order 
of approximation;
in fact such equations contain 
the energy density of
the ultra relativistic matter, which is known only
in the first order (the higher one of the
Einstein equations). Therefore 
 it makes no sense to take into account 
 higher order
 contributions, coming from 
 those equations. 

As soon 
as $(t_r-t)$ is sufficiently large,
it can be easily checked that the solution 
here constructed is completely 
self-consistent 
to the all calculated orders 
of approximation 
in time and contains one physically
arbitrary function of the spatial 
coordinates, $\theta (x^{\gamma })$ which, 
indeed, being a three-scalar, is not 
affected by spatial coordinate 
transformations.  \\
In particular, the terms quadratic in the 
spatial gradients of the scalar field 
are of order  
\beq
\label{ord}
\left(\partial_\alpha\phi\right) ^2 
\approx 
\mathcal{O}\left( \frac{\eta ^2}{a^2}
\frac{1}{\left(t_r-t
\right) ^3} \right)
\eeq
and therefore can be neglected with 
respect to all the inhomogeneous ones. \\
Such solution fails when $t$ approaches
$t_r$ and therefore its validity requires 
that the de-Sitter phase ends (with the
fall of the scalar field in the true 
potential vacuum) when $t$ is yet much 
smaller than $t_r$
 (see below).

\section{Physical considerations}

The peculiar feature of the solution above 
constructed lies 
in the independence of the function 
$\theta$ which, from a 
cosmological point of view, implies the 
existence of a quasi-isotropic inflationary 
solution in correspondence to an arbitrary 
spatial distribution of ultra relativistic 
matter and of the scalar field. 

We get an inflationary picture 
from which the 
Universe outcomes with the appropriate 
standard features, but in presence of a 
suitable spectrum of {\it classical} 
perturbations as due to the small 
inhomogeneities which are 
modelizable according 
to an Harrison--Zeldovic spectrum; in 
fact, expanding the function $\theta$ in 
Fourier series as 
\beq
\label{tet}
\theta(x^\gamma) = 
\frac{1}{(2\pi)^3}\int^{+\infty}_{-\infty}
{\tilde{\theta}\left(\vec{k}\right)
e^{i \vec{k}\cdot\vec{x}}d^3k} \, ,
\eeq
we can impose an 
Harrison--Zeldovic spectrum 
by requiring 
\beq
\label{hz}
{\mid {\tilde{\theta}}\mid} ^2= 
\frac{Z}{\mid k\mid^{3/2}}\, , \qquad Z={\rm const.} \, .
\eeq

However, the following three points 
have to be taken into account to give 
a complete picture for our analysis:

%--------------------list---------
\newcounter{bean}
\begin{list}{(\roman{bean})}{\usecounter{bean}}%
\item 
        limiting (as usual) our attention 
        to the leading order, the validity of 
        the slow-rolling regime is ensured
        by the natural conditions

        \begin{equation}
        \mathcal{O}\left(\sqrt{\chi \Lambda }(t - t_r)\right)\ll 1 \, ,
        \qquad
 \lambda \gg \mathcal{O}(\chi ^2\Lambda )
        \, , 
        \label{slr1}
        \end{equation}

        which respectively translate (\ref{infla2}) 
        and (\ref{infla1}); 
\item 
        denoting by $t_i$ and $t_f$ respectively 
        the beginning and
 the end of the de-Sitter 
        phase, we should have $t_r\gg t_f$
        and the validity of our solution is 
        guaranteed if
        \newcounter{bean2}
        \begin{list}{\alph{bean2}}%{\usecounter{bean2}}%
        \item 
                (a) the flatness of the potential is preserved, 
                i.e.
 $\lambda \phi ^4 \ll \Lambda $: such a 
                requirement coincides, 
as it should, with the 
                second of inequalities (\ref{slr1});
        \item 
                (b) given $\Delta$ as the width of the
                flat region of the potential,
 we require that the 
                de-Sitter phase ends before
                $t$ becomes comparable with $t_r$, i.e.

                \begin{equation}
                \phi (t_f) - \phi (t_i) \sim
                \sqrt{\frac{\sqrt{\chi \Lambda }}{\lambda }}
                \frac{t_f - t_i}{t_r^{3/2}} \sim \mathcal{O}(\Delta )
                \, , 
                \label{slr2}
                \end{equation}

                where we expanded the solution at the first 
                order in
 $t_{i,f}/t_r$; via the usual position
                $(t_f - t_i) \sim 
                \mathcal{O}(10^2)/\sqrt{\chi \Lambda }$, 
                the relation (\ref{slr2}) becomes a constraint 
                for the
 integration constant $t_r$.
        \end{list}
\item 
In order to get an inflationary scenario, able 
to overcome the 
shortcomings present in the 
Standard Cosmological Model, 
we need an exponential 
expansion sufficiently strong. 
For instance we have to require that a region 
of space, corresponding to a cosmological
horizon  $\mathcal{O}(10^{-24}cm)$ when the 
de-Sitter phase starts, now covers all the
actual Hubble horizon $\mathcal{O}(10^{26}cm)$;
the redshift at the end of the de-Sitter phase 
is
 $z \sim \mathcal{O}(10^{24})$, then we should
require $a_f/a_i\sim e^{60}\sim \mathcal{O}(10^{26})$.
Let's estimate the densities perturbations 
(inhomogeneities)
 at the (matter-radiation) 
decoupling age
 ($z\sim \mathcal{O}(10^{4})$) as
$\delta _{in}\sim \mathcal{O}(10^{-4})$; 
if we start by this
same value at the beginning  of inflation
($\delta ^i_{in}$),
 we arrive at the end with
$\delta ^f_{in} \sim (\eta _f/\eta _i)\delta ^i_{in} \sim
\mathcal{O}(10^{-100})$. 
Though these inhomogeneities
 increase as $z^2$ once 
they are at scale greater than the
 horizon, 
nevertheless they reach only
 $\mathcal{O}(10^{-60})$
at the decoupling age.  \\
This result provides support to the
idea that the spectrum of inhomogeneous perturbations cannot
have a classical origin in presence of an inflationary scenario.

\end{list}

%%%%%%%%%%%%%%%%%%%%%%%%%%%%ultima addenda%%%%%%%%%%%%%%%%%%%%%%%%%%%%%

In the considerations above developed,
we regard the ratio of the inhomogeneous terms 
$\epsilon_f$ and $\epsilon_i$ as the
quantity $\delta \rho $ defined in Section 2 and
now we show how this assumption is
(roughly) correct: after the reheating
the Universe is dominated by a homogeneous
(apart from the quantum fluctuations)
relativistic energy density $\rho _r$
to which is superimposed the relic $\epsilon _f$
after inflation; therefore we have
\begin{equation}
\delta \rho = \frac{\epsilon _f}{\rho _r} =
\frac{\epsilon _f}{\epsilon _i}
\frac{\epsilon _i}{\rho _r} =
\left(\frac{a_i}{a_f}\right)^4
\frac{\epsilon _i}{\rho _r} 
\, .
\label{axx}
\end{equation}
Hence our statement follows as soon as we observe
that the inhomogeneous relativistic energy density
before the inflation $\epsilon_i$ and the uniform 
one $\rho _r$, generated by the reheating process, 
differ by only some orders of magnitude.

%%%%%%%%%%%%%%%%%%%%%%%%%%%%%%%%%%%%%%%%%%%%%%%%%
%%%%%%%%%%%%%%%%%%%%%%%%%%%%%%%%%%%%%%%%%%%%%%%%%
%\acknowledgements
\section*{Acknowledgments}

We would like to thank Massimiliamo Lattanzi for his
valuable comments on the problem of cosmological density
perturbations. \\

One of us, G. Imponente, was partly supported 
by {\it ``Progetto Giovani Ricercatori''} within 
the ``Federico II'' University of Naples research projects.

%
%\small


\begin{thebibliography}{99} 




\bibitem{G81}  A. H. Guth, {\it Phys. Rev. D}, (1981) \textbf{23}, 347.

\bibitem{L82}  A. D. Linde, \textit{Phys. Lett.}, (1982) \textbf{108B}, 389.

\bibitem{L83}  A. D. Linde, \textit{Phys. Lett}., (1983) \textbf{129B}, 177.


\bibitem{ST87}  J. Silk and M. S. Turner, \textit{Phys. Rev. D}, (1987) 
\textbf{35}, 419.

\bibitem{CW73}  S. Coleman and E. Weinberg, \textit{Phys. Rev. D}, (1973) \textbf{7}, 1888.

\bibitem{KT90} E.W. Kolb and M.S. Turner, \textit{The Early Universe},
(1990) (Adison-Wesley, Reading).

\bibitem{STA80} A.A. Starobinsky, {\it Phys. Lett. B}, (1980) {\bf 91}, 99.
\bibitem{STA79} A.A. Starobinsky, {\it JETP Lett.} , (1979) {\bf 30}, 682.

\bibitem{dB01} P. de Bernardis et al., {\it Astrophys.J.} (2002) {\bf 564}, 559, (available astro-ph/0105296).

\bibitem{KM02}
A.A. Kirillov and G. Montani, {\it Phys. Rev. D}, (2002) {\bf 66}, 064010.


\bibitem{Star83}  A.A. Starobinsky, \textit{Pis'ma Zh.Eksp.Teor.Fiz.} (1983) \textbf{%
\ 37}, 55.


\bibitem{BKL70}  V.A. Belinskii, I.M. Khalatnikov 
and E.M. Lifshitz, \textit{Adv.\ Phys.} (1970) \textbf{19},  525.

\bibitem{BKL82}  V.A. Belinskii, I.M. Khalatnikov and E.M. Lifshitz, \textit{\
Adv. Phys.} (1982) \textbf{31},  639.

\bibitem{M69}  C.W. Misner, \textit{Phys. Rev. Lett}. (1969) \textbf{22}, 1071.

\bibitem{K93} A.A. Kirillov, {\it Zh.Eksp.Teor.Fiz.} (1993) 
\textbf{103}, 721.
[Sov. Phys. JETP \textbf{76}, 355 (1993)].


\bibitem{KM97} 
A.A. Kirillov and G. Montani 
{\it Phys. Rev. D}, (1997) {\bf 56}, n. 10, 6225. 


\bibitem{IM01}  G. P. Imponente and G. Montani, \textit{Phys. Rev. D},
(2001) \textbf{63}, 103501.

\bibitem{IM02}
G.P. Imponente and G. Montani
 {\it Int. Journ. Mod. Phys. D}, (2002) {\bf 11}, n.8, 1321, 
(available gr-qc/0106028).


\bibitem{LK63}  E.M. Lifshitz and I.M. Khalatnikov, \textit{Adv. Phys.}, (1963) 
\textbf{12},  185.

\bibitem{DT94} K.Tomita and N. Deruelle, \textit{Phys.Rev.D}, 
(1994) {\bf 50}, n.12, 7216.

\bibitem{KHA02} I.M. Khalatnikov et al., {\it CQG},(2002) {\bf 19} 3845 (gr-qc/0204045).
\bibitem{KHA03} I.M. Khalatnikov et al., {\it JCAP}, (2003) {\bf 0303}, 001 (gr-qc/0301119).




\bibitem{M99}  G. Montani, \textit{Class. and Quantum Grav.}, (1999) 
\textbf{16}, 723.


\bibitem{M00}  G. Montani, \textit{Class. and Quantum Grav.}, (2000) 
\textbf{17}, 2205.


\bibitem{MB95}  C. P. Ma and E. Bertschinger, \textit{Astrophys. J.},
(1995) \textbf{455}, 7.



\bibitem{BK73}  V.A. Belinski and I.M. Khalatnikov, \textit{Sov. Phys. JETP},  (1973)
\textbf{36}, 591.

\bibitem{B99}  B.K. Berger, {\it Phys.Rev.D}, (2000) {\bf 61}, 023508 (available gr-qc/9907083).

\bibitem{STA96} D. Polarski, A.A. Starobinsky, {\it CQG}, (1996) {bf 13}, 377.



\end{thebibliography}
\end{document}